\documentclass[apj]{emulateapj}
\usepackage{times}
\usepackage{graphicx}
\usepackage{amssymb}
\usepackage{multirow}
\usepackage{color}
\usepackage{wrapfig}
\usepackage{float}

\usepackage[flushleft]{threeparttable}

\usepackage{lineno}
\usepackage{amsmath}
\usepackage{url}

\newcommand {\fermi }{{\it Fermi}}
\newcommand {\prs } {phase-resolved spectra}
\newcommand {\pas } {phase-averaged spectrum}
\newcommand {\lc } {light curve}
\newcommand {\lcs } {light curves}
\newcommand {\AG } {annular gap}
\newcommand {\CG } {core gap}
\newcommand {\gr } {$\gamma$-ray}
\newcommand {\CR } {curvature radiation}
\newcommand {\SR } {synchrotron radiation}
\newcommand {\pps } {primary particles}
\newcommand {\sps } {secondary particles}
\newcommand {\ef } {electric field}

\newif\ifAMStwofonts
\AMStwofontstrue

\makeatletter

\newcommand{\Rmnum}[1]{\expandafter\@slowromancap\romannumeral #1@}
\makeatother

\shorttitle{The Annular Gap Model for the Crab Pulsar}

\shortauthors{Du et al. 2011}

\begin{document}




\title{radio-to-TeV phase-resolved emission from the Crab pulsar:
  \\ the annular gap model}

\author{Y.~ J.~ Du\altaffilmark{1},~
G.~ J.~ Qiao\altaffilmark{2}~ 
and~
W.~ Wang\altaffilmark{1} } 

\altaffiltext{1}{National Astronomical Observatories, Chinese Academy
  of Sciences, Jia 20 Datun Road, Beijing 100012, China; dyj@nao.cas.cn}
\altaffiltext{2}{School of Physics, Peking University, Beijing 100871,
  China}


\begin{abstract}
The Crab pulsar is a quite young famous pulsar which radiates
multi-wavelength pulsed photons. The latest detection of GeV and TeV
pulsed emission with unprecedented signal-to-noise ratio, supplied by
the powerful telescopes: {\it Fermi}, MAGIC and VERITAS, challenges
the current popular pulsar models, which can be a valuable
discriminator to justify the pulsar high-energy-emission models.

Our work is divided into two steps. First of all, taking reasonable
parameters (the magnetic inclination angle $\alpha=45^\circ$ and the
view angle $\zeta=63^\circ$), we use the latest high-energy data to
calculate radio, X-ray, $\gamma$-ray and TeV light curves from a
geometric view to obtain some crucial information on emission
locations. Secondly, we calculate the phase-averaged spectrum and
phase-resolved spectra for the Crab pulsar and take a theoretical
justification from a physical view for the emission properies as found
in the first step. 
It is found that a Gaussian emissivity distribution with the peak
emission near the null charge surface in the so-called annular gap
region gives the best modeled light curves. The pulsed emission of
radio, X-ray, $\gamma$-ray and TeV are mainly generated from the
emission of {\pps} or {\sps} with different emission mechanisms in the
nearly similar region of the {\AG} located in the only one magnetic
pole, which leads to the nearly ``phase-aligned'' multi-wavelength
{\lcs}. The emission of peak 1 (P1) and peak 2 (P2) is originated from
the annular gap region near the null charge surface, while the
emission of bridge is mainly originated from the core gap region.

The charged particles cannot corotate with the pulsar and escape from
the magnetosphere, which determines the original flowing primary
particles. The acceleration {\ef} and potential in the {\AG} and {\CG}
are huge enough in the several tens of neutron star radii. Thus the
{\pps} are accelerated to ultra-relativistic energies, and produce
numerous {\sps} (pairs) in the inner region of the {\AG} and {\CG}. We
emphasize that there are mainly two types of pairs, i.e., one is
curvature-radiation induced (CR-induced), and the other is
inverse-Compton-scattering induced (ICS-induced).
The {\pas} and {\prs} from soft X-ray to TeV band are produced by four
components: {\SR} from CR-induced and ICS-induced pairs dominates the
X-ray band to soft {\gr} band (100\,eV to 10\,MeV); {\CR} and {\SR}
from the {\pps} mainly contribute to {\gr} band (10\,MeV to $\sim$
20\,GeV); ICS from the pairs significantly contributes to the TeV
{\gr} band ($\sim 20$\,GeV to 400\,GeV). 

The multi-wavelength pulsed emission from the Crab pulsar can be well
modeled with the annular gap and core gap model. To distinguish our
single magnetic pole model from two-pole models, the convincing values
of the magnetic inclination angle and the viewing angle will play a
key role.  

\end{abstract}


\section{Introduction}

The Crab pulsar (PSR B0531+21 or PSR J0534+2200) is the second most
energetic pulsar to date with a spin-down luminosity $\dot{E}_{\rm
  rot}=4.6\times 10^{38} \, \rm erg \, s^{-1}$. It is at a distance of
$d \sim 2$\,kpc, and located in the Crab Nebulae which is a
center-filled remnant of a supernova discovered by the Chinese
``astronomers'' in 1054 AD. The Crab pulsar, a very young pulsar with
a characteristic age $\tau=1240$\,yr and a characteristic magnetic
field of $B_0=3.715\times 10^{12}$\,G, has a spin period of $P=33$\,ms
\citep{2005AJ....129.1993M}.

The Crab pulsar radiates multi-wavelength pulsed emission from radio
($10^{-6}$\,eV) to $\gamma$-ray (up to TeV) band, especially the new
{\gr} results of the sensitive Large Area Telescope (LAT) on board the
{\it Fermi Gamma-ray Space Telescope} ({\fermi}) and a discovery of
TeV pulsed emission by the powerful VERITAS array of atmospheric
Cherenkov telescopes.
{\fermi} LAT presented the high-quality {\gr} (100\,MeV to 20\,GeV)
{\lcs} and spectral data using 8 months of survey data, and predicted
an exponential power-law spectra with cut-off energies of a few GeV
\citep{fermi-Crab}.
To verify the exponential power-law cut-off spectra, the 25 $-$
100\,GeV pulsed emission from the Crab pulsar has been precisely
measured by MAGIC telescope \citep{magic}. It is shown that the
observed cut-off spectrum has a large deviation from the inferred
exponential one.
\cite{veritas} reported that the pulsed emission above 100\,GeV from
the Crab pulsar has been detected by the VERITAS, and showed their
{\lcs} and spectral data with excellent signal-to-noise ratio. They
declared that current popular pulsar models (e.g. outer gap and slot
gap models) cannot explain the detection, and the observation might
not be explained by the {\CR} as the origin of the observed emission
above 100\,GeV.
These findings enable us to obtain considerable insights of the
magnetosphere physics, e.g., acceleration {\ef}, emission region and
the relevant emission mechanisms. The multi-wavelength {\pas} and
{\prs} can discriminate the various pulsar non-thermal emission
models.

There are four major physical or geometrical magnetospheric models
which have previously been proposed to explain pulsed $\gamma$-ray
emission of pulsars: the polar cap model \citep{1994ApJ...429..325D,
  1996ApJ...458..278D}, the outer gap model
\citep{1986ApJ...300..500C, 1986ApJ...300..522C, 1995ApJ...438..314R,
  1997ApJ...487..370Z, 2000ApJ...537..964C, 2004ApJ...604..317Z,
  2007ApJ...666.1165Z, 2008ApJ...688L..25H, 2008ApJ...676..562T,
  2009ApJ...699.1711L}, the two-pole caustic (TPC) model or the slot
gap model \citep{2003ApJ...598.1201D, 2003ApJ...588..430M,
  2004ApJ...606.1143M, 2008ApJ...680.1378H}, and the annular gap model
\citep{2004ApJ...606L..49Q, 2004ApJ...616L.127Q, 2007ChJAA...7..496Q,
  DQHLX+10, Du11}. The distinguishing features of these pulsar models
are different acceleration {\ef} region for primary particles and
relevant emission mechanisms to radiate high energy photons
\citep{Du11}. One of the key discrepancy of the mentioned emission
models is the two important geometry parameters: the magnetic
inclination angle $\alpha$ and the view angle $\zeta$. 

\citet{2000ApJ...537..964C} used a 3D single-pole outer gap model to
present incipient results of {\lcs}, {\pas} and {\prs} for the Crab
pulsar. 
%
%
Eight years later, \citet{2008ApJ...676..562T} improved the outer gap
model and used a modified outer gap model considering the emission
from both poles to calculate {\lcs}, {\pas} and {\prs} from 100\,eV to
10 GeV. 
%
In the case of the larger viewing angle $\zeta=78^\circ$ or
$\zeta=83^\circ$ and intermediate inclination angle $\alpha=45^\circ$,
\cite{2009ApJ...707L.169Z, 2010ApJ...725.2225L} improved the 3D
two-pole outer gap model \citep{2008ApJ...676..562T} to present their
best results of {\lcs}, {\pas} and GeV {\prs}.  They also used
physical emissivities to calculate the {\lcs}, showed the best-fit
{\pas} using emission components from the both magnetic poles, and
indicated that ICS from pairs mainly contributed to the $10^4 -
10^6$\,keV band.
\cite{2008ApJ...680.1378H} used a 3D slot gap model developed from
the TPC model to calculate optical-to-{\gr} {\lcs}, {\pas} and {\prs}
for the Crab pulsar with assuming a broken power-law pair energy
spectrum.
%
%
\cite{2008ApJ...688L..25H} demonstrated that the slot gap model
reproduces at most 20\% of the observed GeV fluxes owing to the small
trans-field thickness.

To well study pulsed emission from pulsars, there may be two methods:
(1) one can use a physical model with a realistic accelerating
electric field to directly calculate the light curves and spectra; (2)
one can also utilize a reasonable assumption of numerical emissivity
to calculate light curves from a geometric point of view; then obtain
some valuable information of radiation locations, finally calculate
the spectra and take a consistent theoretical justification from a
physical point of view. In conventional cases, some scientists chose
the first method. Here we choose the second way to study pulsed
emission from the Crab pulsar.
In this paper, we study multi-wavelength light curves, {\pas} and
{\prs} of the Crab pulsar in the {\AG} model. In \S\,2, we briefly
introduce the annular gap and core gap, pair production and calculate
the acceleration potential in the annular gap. In \S\,3, we model the
multi-wavelength light curves using the annular gap model together
with a core gap. The radio emission region is identified and the
reason for phase-aligned peaks of multi-wavelength {\lcs} is
explained. To model spectra of the Crab pulsar, we also calculate the
multi-wavelength phase-averaged and phase-resolved spectra of
radiation from both primary particles and pairs. Finally, conclusions
and discussions are presented in \S\,4.

\section{The annular gap and core gap}

\subsection{Formation of the Annular Gap and the Core Gap}

\begin{figure}[bt]
\centering
\includegraphics[angle=0,scale=.58]{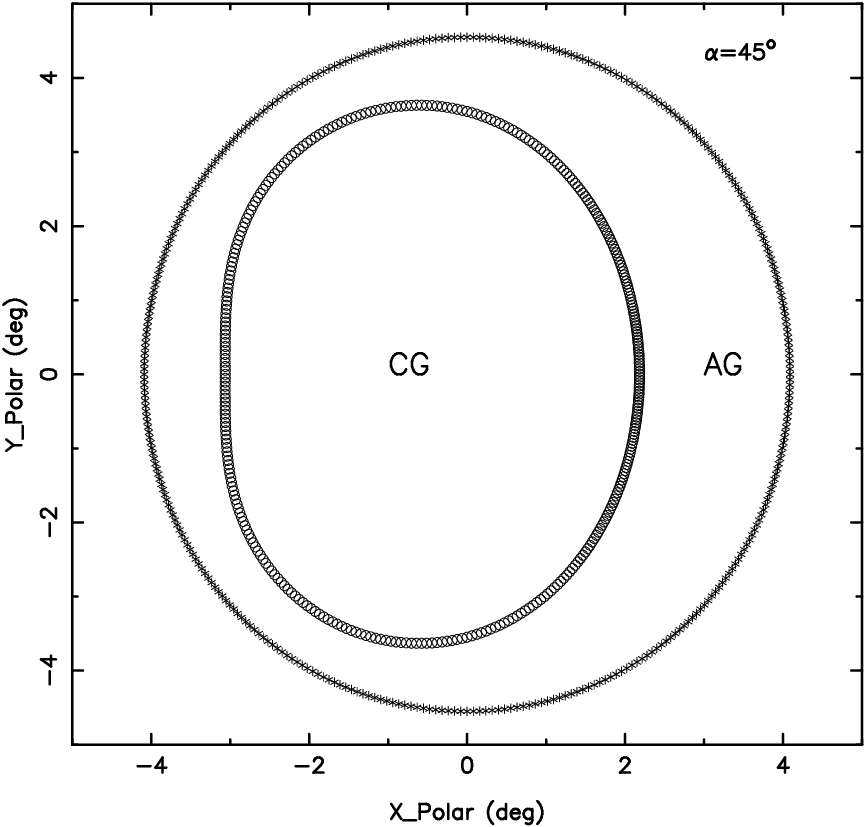}
\caption{The shapes of the annular gap (AG) and core gap (CG) in the
  polar cap region for the crab pulsar with an inclination angle of
  $\alpha=45^\circ$. The core gap region is defined by the footpoints
  of critical field lines on the polar cap, and the annular gap region
  is between the footpoints of critical field lines and last open
  field lines. The annular region becomes more symmetric if $\alpha$
  is smaller.}
\label{POL}
\end{figure}

As noted by \cite{Du11}, the open-field-line region of a pulsar
magnetosphere is divided into two isolated parts by the critical field
lines which denote a set of special field lines that satisfy the
condition of $\mathbf{\Omega \cdot B}=0$ at the light cylinder. The
core region around the magnetic axis is defined by the critical field
lines, and the annular region is located between the critical field
lines and the last open field lines (see Figure~\ref{POL}). The width
of the annular polar region is anti-correlated with the pulsar period,
it is therefore larger for pulsars with smaller spin periods. The
annular acceleration potential is negligible for older long-period
pulsars, but very important for pulsars with a small period, e.g.,
millisecond pulsars and young pulsars. The acceleration {\ef} extends
from the pulsar surface to the null charge surface or even beyond
it. The annular gap has a sufficient thickness of trans-field lines
and a wide altitude range for particle acceleration. In the annular
gap model, the high energy emission is generated in the vicinity of
the null charge surface \citep{DQHLX+10}. This leads to a wide
$\gamma$-ray emission beam \citep{2007ChJAA...7..496Q}. The radiation
components from both the core gap and the annular gap can be observed
simultaneously by one observer \citep{2004ApJ...616L.127Q} if the
inclination angle and the viewing angle are suitable.

\subsection{Acceleration Potential in the Annular Gap}

Here we will explore the formation mechanism of the acceleration
electric field in the annular gap. To give a simplified picture for
the powerful acceleration electric field along an open field line, we
attempt to derive 1D continuous solution for the acceleration
potential. Then we use this realistic acceleration field to obtain the
Lorentz factor $\gamma_{\rm p}$ of the primary particle which is from
the balance of the acceleration and curvature radiation reaction (see
Equation (\ref{gam_p}) in \S2.3). The $\gamma_{\rm p}$ is not a true
maximum Lorentz factor but a valuable criterion, it can be regarded as
an upper limit of the real maximum Lorentz factor of the primary
particles which shape the obsered spectra. 

Under the assumption of a fully charge-separated magnetosphere, the
outer gap can be formed \citep{1986ApJ...300..500C}. However, if
abundant pairs are produced, this lead to a large neutral charge
component in the magnetosphere. In this case, the acceleration
mechanism of particles becomes different from the case of outer gap
model.

The pulsar magnetosphere is filled with charge-separated pair plasma,
and the charged particles can not co-rotate with the neutron star near
the light cylinder and must escape from the magnetosphere. This is the
generation mechanism for acceleration {\ef}. 
The annular gap and the core gap simultaneously export charged
particles with opposite sign, which can lead to the circuit closure in
the whole magnetosphere. The parallel acceleration electric field
($E_{\parallel}$) in the annular gap and core gap regions are
opposite. As a result, $E_{\parallel}$ vanishes at the boundary (the
critical field lines) between the annular and the core regions and
also vanishes along the closed field lines. The positive and the
negative charges are accelerated from the core and the annular
regions, respectively.

\begin{figure}[tb]
\centering
\includegraphics[angle= 0,scale=.58]{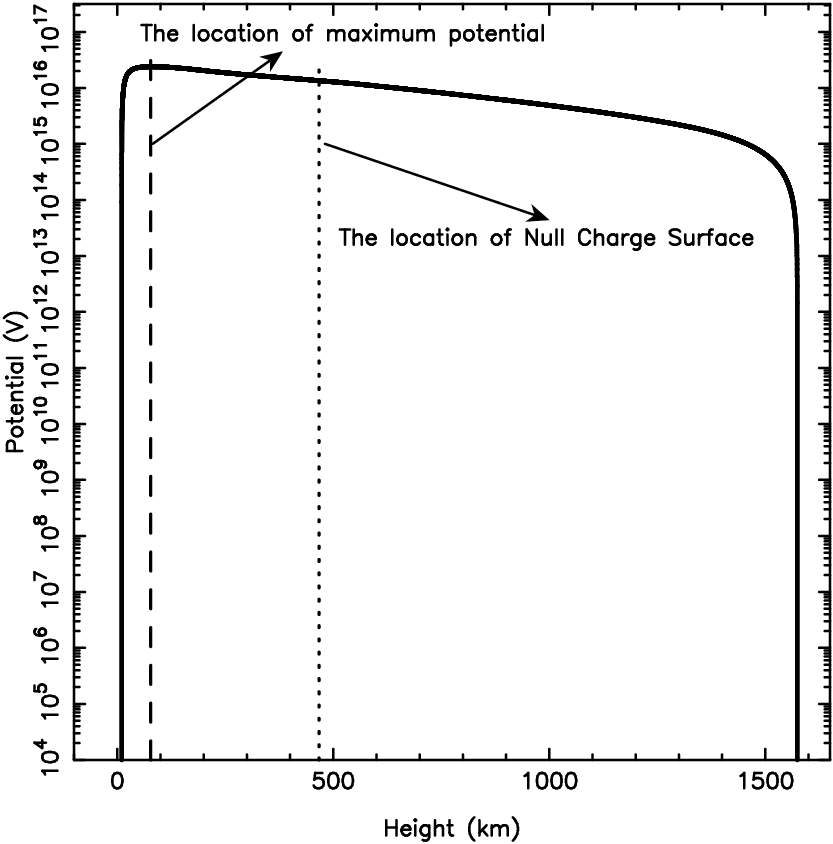}
\caption{Acceleration potential on an open field line with a magnetic
  azimuthal $\Psi=0^\circ$ in the annular gap. The locations of
  maximum potential and null charge surface are marked. The maximum
  potential drop on this field line is comparable to the one generated
  by the unipolar effect. } \label{POT}
\end{figure}

To unveil the acceleration potential in the {\AG} region, we now
consider a tiny magnetic tube embedded in an open field line. We
assume that the particles flow out of the co-rotating magnetosphere at
a radial distance about $r_{\rm out}\sim R_{\rm LC}=1.57\times
10^3$~km, and that the charge density of flowing-out particles
$\rho_{\rm b}(r_{\rm out})$ is equal to the local Goldreich-Julian
(GJ) charge density $\rho_{\rm gj}(r_{\rm out})$ \citep{GJ69}. For any
altitudes $r<r_{\rm out}$,
$\rho_{\rm b}(r) < \rho_{\rm gj}(r)$. 
The acceleration electric field therefore exists along the field line,
and cannot vanish until approaching the altitude of $r_{\rm out}$.

In a static dipole magnetic field configuration, the field
components can be described as
$\mathbf{B}_r=\frac{2\mu \cos\theta}{r^3}\mathbf{n}_r$
and 
$\mathbf{B}_{\theta}=\frac{\mu \sin\theta}{r^3}\mathbf{n}_{\theta}$, 
here $\theta$ is the zenith angle in magnetic coordinate, and $B_0$ is
the surface magnetic field. Thus the magnetic field strength at a
altitude $r$ is
$B(r)=\frac{B_0 R^3}{2}\frac{\sqrt{3\cos^2\theta+1}}{r^3}$. 

In the co-rotating frame, the equation for acceleration potential
$\Psi$ is
%
%
%
\begin{equation}
\nabla^2 \Psi = -4\pi(\rho_{\rm b} - \rho_{\rm gj}). \label{acp}
\end{equation}
Although the same Possion equation for acceleration field is used by
all of pulsar emission models (some models also considered the
relativistic effect), the key discrepancy among these models is how to
obtain and explain the difference between flowing charge density and
the local GJ density ($\rho_{\rm b}-\rho_{\rm gj}$). Normally it is
assumed that $\rho_{\rm b}-\rho_{\rm gj} \sim 0$ at the star surface.
This leads to the quite different results of the acceleration electric
field.

Using the conservation laws of the particle number and magnetic flux
in the magnetic flux tube, the difference between the flowing charge
density and local GJ charge density at the altitude $r$ can be written
as
\begin{equation}
\rho_{\rm b}(r)-\rho_{\rm gj}(r) =
-\frac{\Omega B(r)}{2 \pi c}(\cos\zeta_{\rm out}-\cos\zeta), 
\label{rho}
\end{equation}
where $\Omega=2\pi/P$ is the angular velocity, $P$ is the rotation
period, and $\zeta$ and $\zeta_{\rm out}$ are the angle between the
rotational axis and the $B$ field direction at $r$ and $r_{\rm out}$,
respectively. Wang et al. (2006) found
\begin{equation}
\cos\zeta
=\cos\alpha\cos\theta_{\mu}-\sin\alpha\sin\theta_{\mu}\cos\psi,  
\label{zet}
\end{equation}
where $\psi$ and $\theta_{\mu}$ are the azimuthal angle and the
tangent angle (half beam angle) in the magnetic field coordinate,
respectively. Combine Equations (\ref{acp}), (\ref{rho}), and
(\ref{zet}), we then obtain one-dimensional two-order differential
equation for the acceleration potential, i.e.
\begin{equation}
\frac{\mathrm d^2 \Psi}{\rm d \theta^2} - \frac{s^{''}_{\rm
    \theta}}{s^{'}_{\rm \theta}}\frac{\mathrm d \Psi}{\mathrm d
  \theta} = \frac{\Omega B_0 R^3}{c r^3}{\sqrt{3\cos^2\theta+1}}
(\cos\zeta_{\rm out}-\cos\zeta){s^{'}_{\rm \theta}}^2,
\label{eee}
\end{equation}
where $s^{'}_{\rm \theta}$, $s^{''}_{\rm \theta}$ are given by
\begin{displaymath}
s^{'}_{\rm \theta}=\frac{\mathrm{d} s}{\mathrm{d}
  \theta}=\sqrt{r^2+\left (\frac{\mathrm{d} r}{\mathrm{d} \theta}
  \right )^2},
\end{displaymath}
\begin{displaymath}
s^{''}_{\rm \theta}=\frac{\mathrm{d} s^{'}_{\rm \theta}}{\mathrm{d} \theta}, 
\end{displaymath}
\begin{displaymath}
r=R_{\rm e}\sin^2{\theta},
\end{displaymath}
here $R_{\rm e}$ is a field line constant, which denotes for the
maximum length of a certain point on a field line. Then substituting
$\tan\theta_{\mu}=\frac{3\tan\theta}{2-\tan^2\theta}$
\citep{1998A&A...333..172Q} into Equations (\ref{zet}) and
(\ref{eee}), we can solve the Equation (\ref{eee}) , and achieve the
1-D solution for the electric potential $\Psi$ along a magnetic filed
line with a magnetic azimuthal of $\psi=0^\circ$ for the Crab pulsar,
as shown in Figure \ref{POT}. The maximum potential drop on this field
line is comparable to the potential
\begin{displaymath}
\Delta V = \frac{\Omega B R^2}{2c}=1.0\times10^{14}B_{12}R_6^2P^{-1},
\end{displaymath}
which is generated by the unipolar effect.  The acceleration potential
is quite huge in the inner region of annular gap, and the {\pps} are
therefore accelerated to ultra-relativistic energy with large Lorentz
factors of $\gamma \sim 10^6 - 10^7$. Simultaneously, the accelerated
{\pps} emit abundant {\gr} photons through ICS and CR process, then
dense ${\rm e}^\pm$ pairs are generated via $\gamma-$B (photon
magnetic absorption) process.


\subsection{Pair Production}

Since ICS and {\CR} are two effective radiation mechanisms to generate
high energy {\gr} photons, ${\rm e}^\pm$ pairs can be generated by
these ICS and CR photons emitted from the accelerated {\pps} in both
annular gap and core gap. Thus three gap modes exist for pair
production, namely, CR gap, thermal ICS gap and resonant ICS gap
\citep{1997ApJ...478..313Z,1997ApJ...491..891Z}, which will be briefly
introduced below.

In the traditional inner gap model \citep{RS75}, $\gamma-B$ process
plays a very important role, two conditions should be satisfied at the
same time for pair production: (1) to produce enough high energy
$\gamma$-ray photons, a strong enough potential drop should be
reached; (2) for pair production, the energy component of $\gamma$-ray
photons perpendicular to the magnetic field must satisfy the condition
of $E_{\gamma,\perp} \geq 2 m_e c^2$.

The accelerated particles are assumed to flow along a field line in a
quasi-steady state. Using the calculated acceleration electric field,
we can obtain the Lorentz factor $\gamma_{\rm p}$ of the primary
particle from the curvature radiation reaction
\begin{equation}
\gamma_{\rm p} = (\frac{3\rho^2 E_{\parallel}}{2e})^{\frac{1}{4}} = 2.36\times
  10^7{\rho_7}^{0.5} E_{\parallel,\, 6}^{0.25}, 
\label{gam_p}
\end{equation}
where $e$ is the charge of an electron, $\rho_7$ is the curvature
radius in units of $10^7$\,cm and $E_{\parallel,\, 6}$ is the
acceleration electric field in units of $10^6 \rm V\, cm^{-1}$.

In $\gamma-$B process, the conditions for pair production are that the
mean free path of $\gamma$-ray photon in strong magnetic field is
equal to the gap height, $l\approx h$. The mean free path of
$\gamma$-ray photon is given by \citep{erber66}
\begin{equation}
l=\frac{4.4}{e^2/\hbar c}\frac \hbar {m_ec}\frac{B_c}{B_{\perp
}}\exp (\frac 4{3\chi }),
\end{equation}
where $B_{\rm c}=4.414\times 10^{13}$~G is the
critical magnetic field, $\hbar$ is the reduced Planck's constant,
\begin{equation}
\chi =\frac{E_\gamma }{2m_ec^2}\sin \theta \frac
B{B_{\rm c}}=\frac{E_\gamma }{2m_ec^2}\frac{B_{\perp }}{B_{\rm c}},
\end{equation}
and $B_\perp$ is the magnetic field perpendicular to the moving
direction of $\gamma$ photons, which can be expressed as (RS75)
\begin{equation}
B_{\perp }\approx \frac h\rho B\approx \frac l\rho B.
\end{equation}
Here $l\approx h$ is the condition for gap sparks (pair production) to
take place. $\rho$ is curvature radius of a spot on a magnetic field
line. For a dipole magnetic configuration, it can be estimated as
\begin{equation}
\rho \approx \frac 43(\lambda Rc/\Omega )^{1/2}
\end{equation}
\citep{1997ApJ...491..891Z} if the spot position is near the neutron
star surface. Here $\lambda$ is a parameter to show the field lines,
$\lambda=1$ corresponding to the last open field line. The
characteristic energy from the curvature radiation process can be
written as
\begin{equation}
E_{\gamma,cr}= \hbar \frac {3 \gamma^3 c }{2 \rho}. \label{Ecr}
\end{equation}
Then the CR gap height $h_{\rm CR}$ is 
\begin{equation}
h_{\rm CR} \simeq 5\times 10^3 P^{3/7}B_{12}^{-4/7}\rho_6^{2/7} \rm \; cm
\end{equation}
\citep{1997ApJ...491..891Z}. 

The so-called CR-mode gap implicates that the pair production cascades
are dominated by the {\gr} photons emanated from {\CR} process of the
accelerated {\pps}, this gap is somehow like the RS gap \citep{RS75}.
The resonant ICS gap is formed from the resonant scattering of soft
photons, which is a quantum effect with a large electron scattering
cross section estimated as $\sigma \sim 10^8 \sigma_{\rm T}$
($\sigma_{\rm T}$ is the Thompson cross section) in strong magnetic
fields. \cite{1997ApJ...491..891Z} obtained the gap height ($h_{\rm
  res}$) of the resonant ICS mode
\begin{equation}
h_{\rm res} \simeq 1.1 \times 10^3 P^{1/3}B_{12}^{-1}\rho_6^{1/3} \rm \;cm,
\end{equation}
and the Lorentz factor ($\gamma_{\rm 2,\, res}$) of pairs for the
resonant ICS mode
\begin{equation}
\gamma_{\rm 2,\,res}=890P^{-1/3}B_{12}\rho_6^{2/3}.
\end{equation}
The thermal ICS gap is determined by those thermal-peak photons which
has the maximum photon number density of the Planck spectrum at a
certain temperature.
It has a lower gap height $h_{\rm th}$
\begin{equation}
h_{\rm th} = 2.7\times 10^2
P^{2/5}B_{12}^{-3/5}\rho_6^{1/5}T_6^{-1/5} \rm \; cm,
\end{equation}
but lead to a larger Lorentz factor ($\gamma_{\rm 2,\, th}$) for the
{\sps}
\begin{equation}
\gamma_{\rm 2,\,th} = 3.7\times 10^3
P^{-2/5}B_{12}^{3/5}\rho_6^{4/5}T_6^{1/5}
\end{equation}
\citep{1997ApJ...491..891Z}.

These two CR and ICS gaps, which have relatively lower gap heights,
would dominate the inner gap breakdown
\citep{1997ApJ...491..891Z,smsp}. The pairs can also be abundantly
generated by the {\pps} escaped from the inner gap within a few
neutron star radii, and they could have two major energy distributions
due to the different types of {\pps}. We will therefore use two
different pair energy distributions for CR and ICS pairs to calculate
the {\pas} and {\prs} for the Crab pulsar.
%

\section{Modeling the multi-wavelength light curves and spectra 
for the Crab pulsar}

To explain the multi-wavelength {\lcs} with nearly aligned peak phase
for the Crab pulsar, we should obtain the high signal-to-noise data,
which are adopted from Figure 2 of \citep{fermi-Crab}. We also
reprocessed the {\fermi} {\gr} data to obtain the three {\gr} band
light curves (see left panels of Figure \ref{MLC}) in the following
steps:
\begin{enumerate}
\item Limited by the timing solution for the Crab
pulsar\footnote{http://fermi.gsfc.nasa.gov/ssc/data/access/lat/ephems/}
from the Fermi Science Support Center (FSSC), we reprocessed the
original data observed from 2008 August 4 to 2009 April 8.
\item We selected photons of 0.1-300\,GeV in the ``Diffuse" event class,
within a radius of $2^\circ$ of the Crab pulsar position
(RA$=83.63^\circ$, DEC$=22.01^\circ$) and the zenith angle smaller
than $105^\circ$.
\item As done by \cite{fermi-Crab}, we used ``fselect" to select
  photons of energy $E_{\rm GeV}$ within an angle of
  $<\max\left[\,6.68 - 1.76\log_{10}(E_{\rm GeV}),1.3\,\right]$
  degrees from the pulsar position.
\item We then obtained the rotational phase for each photon using the
  tempo2 \citep{2006MNRAS.369..655H} with the {\fermi} plug-in.
\item Finally we obtained the multi-wavelength $\gamma$-ray light
  curves with 256 bins, as presented in Figure \ref{MLC} (left
  panels). Two sharp phase-aligned peaks have a phase separation of
  $\delta \phi \sim 0.4$.
%
\end{enumerate}

An acceptable model should have reasonable input parameters (e.g.,
magnetic inclination angle $\alpha$ and viewing angle $\zeta$), and
consistently produce multi-wavelength light curves with phase-aligned
peaks and bridge emission and phase-resolved spectra for the Crab
pulsar .

\subsection{Light Curves Modeling}

We adopted the method \citep[see details in \S3.1 of][including basic
  formulae and gap physics]{Du11} to model the {\lcs}. The key idea of
modeling {\lcs} is to project the radiation intensity of every spot on
each open filed line (in either {\AG} or {\CG}) to the
``non-rotating'' sky, with considerations of physical effects. Here
the emissivities are numerically assumed to facilitate calculations,
they are however consistent with the physically calculated spectra, as
noted in Figure 8 of \cite{Du11}.
Some model parameters for both the annular gap and core gap should be
adjusted for the emission regions where the corresponding waveband
emission are generated. The framework of the annular gap model as well
as the details of coordinate system have been presented in
\citet{DQHLX+10}, which can be used for simulation of the
multi-wavelength light curves of the Crab pulsar. We adopted the
inclination angle of $\alpha=45^{\circ}$ and the viewing angle
$\zeta=63^{\circ}$ which were obtained from the {\sl Chandra} X-ray
torus fitting \citep{NG08}. The modeling methods are briefly
delineated as follows.

\begin{figure*}
\centering
\includegraphics[angle=0,scale=.68]{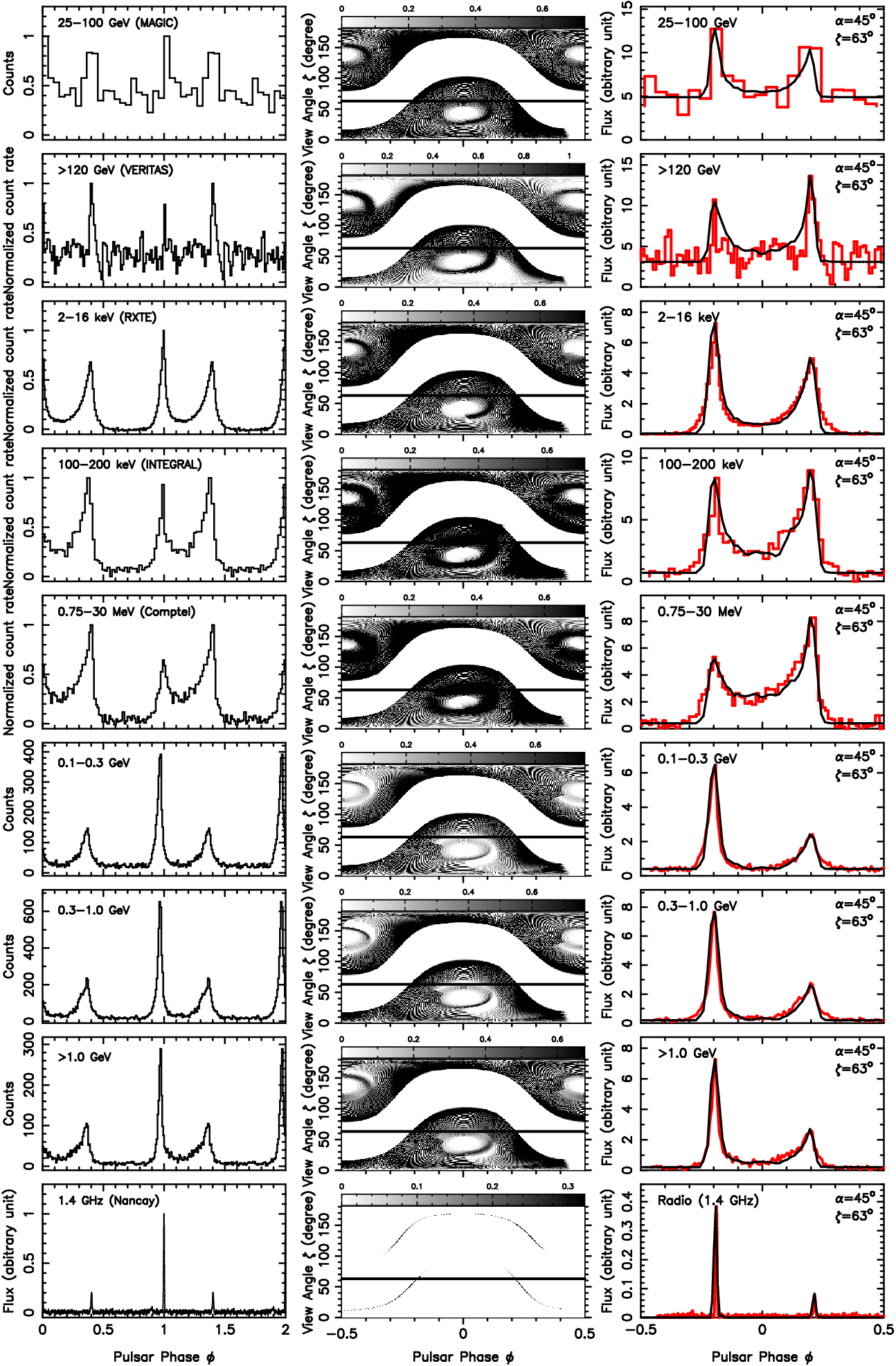}
\caption{Multi-wavelength (radio, X-ray and {\gr}) light curves for
  the Crab pulsar. The observations are shown in the left panels, and
  the data of RXTE, INTEGRAL, Comptel and Nancay are taken from Figure
  2 of \cite{fermi-Crab}. The 25$-$100\,GeV light curve data are taken
  from \cite{magic}, and the TeV ($>$120\,GeV) data are taken from
  \cite{veritas}. The photon sky-map (middle panels) for an
  inclination angle and the corresponding modeled light curves (right
  panels) for a viewing angle of $\zeta=63^\circ$ are also presented,
  using the single-pole {\AG} model. Our {\AG} model can well explain
  the multi-wavelength {\lcs} with phase-aligned peaks.  \\ (A color\,
  version of this figure is available in the online
  journal.)} \label{MLC}
\end{figure*}

\begin{enumerate}
\item We first use the critical field line to separate the polar cap
  region into the annular and core gap regions. Then we use the
  so-called ``open volume coordinates" ($r_{\rm OVC}, \psi_{\rm s}$)
  to label the open field lines of the annular gap and core gap,
  respectively. Here $r_{\rm OVC}$ is the normalized magnetic
  colatitude and $\psi_{\rm s}$ is the magnetic azimuthal. For the
  annular gap, we define the inner rim $r_{\rm OVC,\, AG} \equiv 0$
  for the critical field lines and the outer rim $r_{\rm OVC,\, AG}
  \equiv 1$ for the last open field lines; while for the core gap, We
  define the outer rim $r_{\rm OVC,\, CG} \equiv 1$ for the critical
  field lines and the inner rim $r_{\rm OVC, \, CG} \equiv 0$ for the
  magnetic axis. We also divide both the annular gap ($0 \lesssim
  r_{\rm OVC, \, AG} \lesssim 1$) and the core gap ($0.1 \lesssim
  r_{\rm OVC, \,CG} \lesssim 1$) into 40 rings for calculation.

\item When calculating the emissivities for modeling light curves, we
  postulate that the emissivities along one open field line have a
  Gaussian distribution rather than the frequently used assumption of
  the uniform emissivity along an open field line
  \citep{2003ApJ...598.1201D,2010ApJ...709..605F} for both the annular
  gap and the core gap. To justify this key assumption, we already
  used the realistic acceleration field to plot the flux against
  emission altitude along an open field line for the Vela, as shown in
  Fig. 8 of \cite{Du11}. From that figure, one can clearly see that
  the flux is likely to have a Gaussian distribution against altitude
  near the peak position. The peak emissivities $I_{\rm p}(\theta_{\rm
    P}, \psi_{\rm s})$ may follow another Gaussian distribution
  against $\theta$ for a bunch of open field lines
  \citep{2000ApJ...537..964C, 2003ApJ...598.1201D,
    2010ApJ...709..605F}.  As seen above, we use two different
  Gaussian distributions to describe the emissivities on open field
  lines for both the annular gap and the core gap. The model
  parameters are adjusted to maximally fit the observed
  multi-wavelength light curves.

%

\item To derive the ``photon sky-map" in the observer frame, we first
calculate the emission direction of each emission spot ${\bf n_{\rm
    B}}$ in the magnetic frame; then use a transformation matrix
$T_{\rm \alpha}$ to transform ${\bf n_{\rm B}}$ into ${\bf n_{\rm
    spin}}$ in the spin frame; finally use an aberration matrix to
transform ${\bf n_{\rm spin}}$ to ${\bf n_{\rm observer}} = \{ {\bf
  n_{\rm x}, \, n_{\rm y}, \, n_{\rm z} }\}$ in the observer
frame. Here $\phi_{\rm 0} = \arctan ({\bf n_{\rm y}}/{\bf n_{\rm x}})$
and $\zeta = \arccos({\bf n_{\rm z}}/\sqrt{ {\bf n_{\rm x}}^2 + {\bf
    n_{\rm y}}^2 + {\bf n_{\rm z}}^2})$ are the rotation phase
(without retardation effect) of the emission spot with respect to the
pulsar rotation axis and the viewing angle for a distant, nonrotating
observer. The detailed calculations for the aberration effect can be
found in \cite{LD10}.

\begin{figure}
\centering
\includegraphics[angle=0,scale=.58]{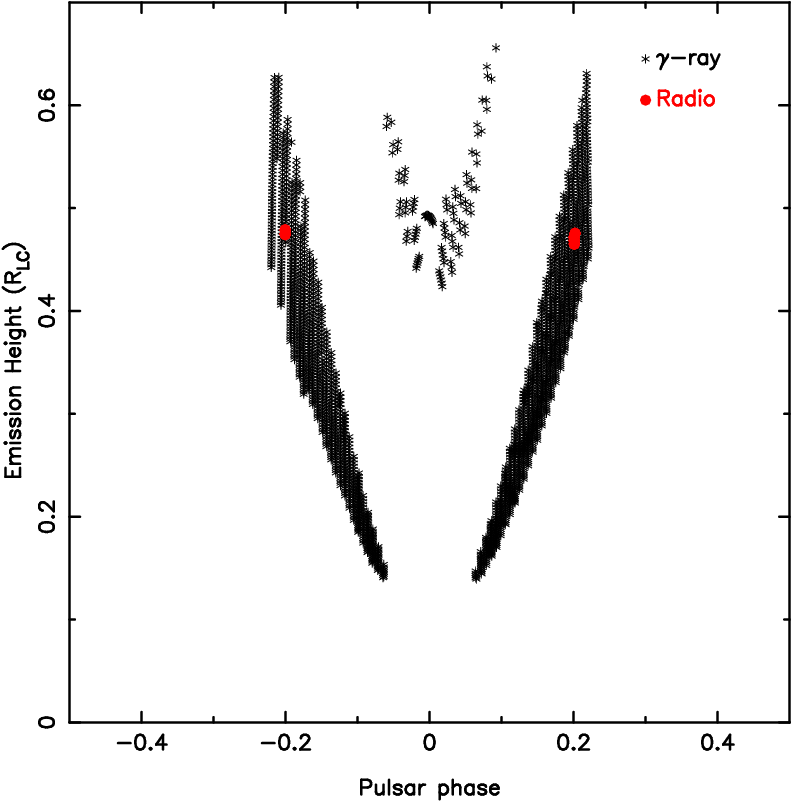}
\caption{{\gr} and radio emission altitudes for the Crab pulsar with a
  viewing angle $\zeta=63^\circ$ in the annular gap model. The {\gr}
  peak and radio (1.4\,GHz) emission altitudes are nearly overlapped
  in the annular gap region, which leads to the phase-aligned peaks of
  the two energy band. While bridge emission are generated from the
  core gap region.\\ (A color version of this figure is available in
  the online journal.)} \label{EMH}
\end{figure}

\item We also add the phase shift $\delta \phi_{\rm ret}$ caused by
  the retardation effect to the emission phase, that is
  $\phi=\phi_{\rm 0}-\delta \phi_{\rm ret}$. There is no minus sign
  for $\phi_{\rm 0}$ because of our coordinate system different from
  other models.

\item The ``photon sky-map'', defined by the binned projected emission
  intensities on the ($\phi$, $\zeta$) plane, can be plotted in 256
  bins (see middle panel of Figure \ref{MLC}). The corresponding light
  curves cut by a line of sight with a viewing angle
  $\zeta=63^{\circ}$ are finally obtained. For the viewing angle
  $\zeta=63^{\circ}$, any magnetic inclination angles of $\alpha$
  between $40^{\circ}$ and $65^{\circ}$ in the annular gap model can
  produce light curves with two peaks and a large peak separation
  similar to the observed ones. The emission from the single pole is
  favored for the Crab pulsar in our model.
\end{enumerate}

The modeled light curves from radio to TeV band are presented in
Figure \ref{MLC} (black solid lines), and the key model parameters are
listed in Table~\ref{tbl_1}. Emission from P1 and P2 of
multi-wavelength {\lcs} originates from the annular gap region in the
vicinity of the null charge surface, while bridge emission comes from
the core gap region. We emphasize that all the multi-wavelength
emission are originated from only one magnetic pole, our {\AG} is
therefore a single-pole magnetospheric model.

\begin{table}[htb]
\raggedright
\caption{Model parameters for multi-wavelength light curves of the
  Crab pulsar
\label{tbl_1}}
{\scriptsize  
\begin{tabular}{lccccccccc} 
\hline \hline
Band & $\kappa$ & $ \lambda$ & $\epsilon$ & $\sigma_{\rm A}$ &
$\sigma_{\rm \theta,\,A}$ & $\sigma_{\rm C}$ & $\sigma_{\rm
  \theta,\,C1}$ & $\sigma_{\rm \theta,\,C2}$ \\
\hline
25--100 GeV & 0.65  & 0.6 & 0.8 & 0.4 & 0.0045 & 0.25  & 0.0023  & 0.0035 \\
$>$120 GeV & 0.68  & 0.85 & 0.8 & 0.3 & 0.0035 & 0.25 & 0.0025  & 0.0032 \\
2--16 keV  & 0.50  & 0.8 & 1.17 & 0.5 & 0.005 & 0.3  & 0.0022  & 0.0035 \\
100-200 keV & 0.50  & 0.8 & 0.8 & 0.6 & 0.006 & 0.35  & 0.0035  & 0.0045 \\
0.75-30 MeV & 0.52 & 0.8 & 0.7 & 0.5 & 0.004 & 0.3  & 0.005  & 0.006 \\
%
0.1--0.3 GeV & 0.53  & 0.8 & 0.6 & 0.5 & 0.004  & 0.15 & 0.0012 & 0.0012 \\
0.3--1.0 GeV & 0.54  & 0.8 & 0.7 & 0.5 & 0.005 & 0.15 & 0.0014 & 0.0014 \\
$>$1.0 GeV & 0.57  & 0.8 & 0.5 & 0.5 & 0.005 & 0.35  & 0.0035 & 0.0035 \\
\hline
\end{tabular}  }

{\bf Note.} $\kappa$ and $\lambda$ are two geometry parameters to
determine the peak altitude in the annular gap; $\epsilon$ is a
parameter for the peak altitude in the core gap; $\sigma_{\rm A}$ and
$\sigma_{\rm C}$ are length scales for the emission region on each
open field line in the annular gap and the core gap in units of
$R_{\rm LC}$, respectively; $\sigma_{\rm \theta,\,A}$ is the transverse
bunch scale for field lines in the annular gap; $\sigma_{\rm
  \theta,\,C1}$ and $\sigma_{\rm \theta,\,C2}$ are the bunch scale for
field lines of $-180^\circ<\psi_{\rm s}<90^\circ$ and
$90^\circ<\psi_{\rm s}<180^\circ$ in the core gap, respectively. The
detailed description of these symbols can be found in \citep{Du11}.


\end{table}

With well-coordinated efforts for pulsar radio timing program,
\cite{fermi-Crab} determined the phase lag between radio emission and
$\gamma$-ray light curves. The first {\gr} peak comes earlier than the
1.4\,GHz radio pulse by a small phase of $\sim 0.00853$, but they are
nearly phase-aligned.
From {\lc} modeling, we can obtain the emission locations for each
band. The result of {\gr} and radio emission altitudes for the Crab
pulsar is shown in Figure \ref{EMH}. The region for the radio emission
is mainly located at a altitude of $\sim 0.5R_{\rm LC}$ on some
certain filed lines, for P1, the magnetic azimuthal $\psi$ is in the
range of $-98^\circ$ to $-96^\circ$, while $106^\circ$ to $109^\circ$
for P2. It is generated in an intermediate-altitude {\AG} region,
which might be due to the coherence condition and propagation effects.
It is found that the positions of both {\gr} (X-ray) peak and radio
peak (1.4\,GHz) are overlapped, this leads to the nearly phase-aligned
pulse peaks of multi-wavelength emission, except for several GHz radio
emission due possibly to plasma propagation effects. Nevertheless, the
$\gamma$-ray emission altitudes are above the lower bound of the
height determined by $\gamma-B$ absorption \citep{LD10}. Based on our
model, not all $\gamma$-ray pulsars can be detected in the radio band,
and not all radio pulsars can have a $\gamma$-ray beam with
sufficiently high flux towards us. The beam shapes and intensities of
$\gamma-$ray and radio can evolve with pulsar ages.






\subsection{Multi-wavelength Spectra for the Crab Pulsar}

In this section, we will use the {\AG} model to calculate the
multi-wavelength {\pas} and {\prs} for the Crab
pulsar. \cite{kuiper01} achieved seven band phase-resolved spectra
(LW1, P1, TW1, Bridge, LW2, P2, TW2) and the phase-averaged spectrum
of the Crab pulsar with the {\sl EGRET} {\gr} data. We will also add
the new high-quality {\fermi}, MAGIC and VERITAS data to the total
{\pas}. The {\gr} emission is believed to be originated from the
curvature radiation of primary particles \citep{2008ApJ...676..562T,
  2008ApJ...680.1378H, Meng08}, which generally gives a
super-exponential power-law spectrum with the cut-off energy of a few
GeV. However, the TeV (20 to 400\,GeV) emission mechanism remains a
mystery, and it requires a global {\pas} fitting to resolve this
problem.

\begin{figure*}
\centering
\includegraphics[angle=-90,scale=.58]{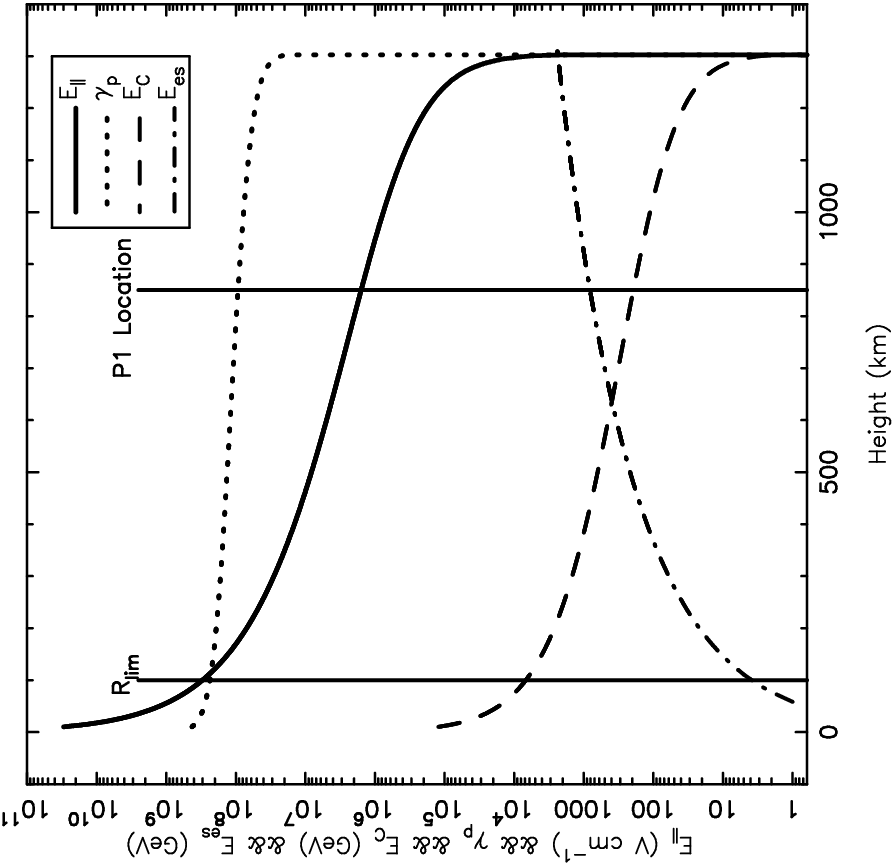}
\includegraphics[angle=-90,scale=.58]{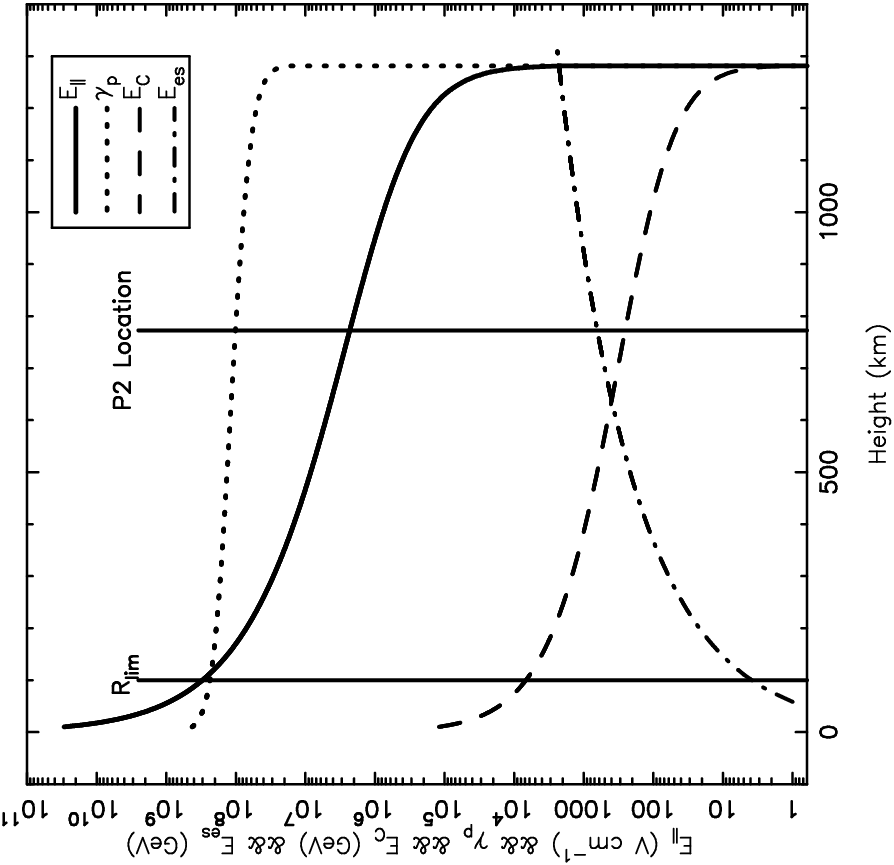}
\caption{Four dynamic parameters along two open field lines where P1
  (top panel) and P2 (bottom panel) are mainly originated. The
  acceleration {\ef} $E_\parallel$ (solid line) is quite huge in the
  inner region of {\AG} below the altitude of $R_{\rm lim}$, which
  results in the generation of numerous pairs. The Lorentz factor of
  primary particles $\gamma_{\rm p}$ (dotted line) is derived from the
  {\CR} reaction. $E_{\rm C}$ (dashed line) denotes for the
  characteristic energy of {\CR} emitted from {\pps}. $E_{\rm es}$
  (dot-dash line) is the escape (maximum) photon energy for a certain
  altitude due to the $\gamma-$B absorption.} \label{DYN}
\end{figure*}

%
We first discuss the particle dynamics in the annular gap. After
exploring the formation mechanism of acceleration {\ef}, the relevant
dynamic parameters: acceleration {\ef} $E_\parallel$, Lorentz factor
of {\pps} $\gamma_{\rm p}$, characteristic energy of {\CR} emitted
from {\pps} $E_{\rm C}$ and escape photon energy $E_{\rm es}$ are
calculated for two field lines where P1 and P2 are mainly
originated. The calculated results are shown in Figure \ref{DYN}.
As introduced in \S\,2.2, the derived acceleration {\ef} $E_\parallel$
is quite huge in the inner region of {\AG} below the altitude of
$R_{\rm lim}$, which leads to the generation of numerous pairs via
$\gamma-$B absorption effect. The Lorentz factor of primary particles
$\gamma_{\rm p}$ is derived from the {\CR} reaction using the
Equation~(\ref{gam_p}). However, the actual Lorentz factor of the
{\pps} is smaller than the one shown in Figure \ref{DYN} if other
energy loss mechanisms (e.g. ICS loss rate) as well as relativistic
and pair screening effects are taken into account.  $E_{\rm C}$
denotes for the characteristic energy of {\CR} emitted from {\pps},
derived by the Equations (\ref{gam_p}) and (\ref{Ecr}). $E_{\rm es}$,
due to the $\gamma-$B absorption effect based on the
Equation~(\ref{tau}), is the escape (maximum) photon energy for a
certain altitude.

We will further consider the emission mechanisms for the Crab
pulsar. The photon number spectrum of the synchrotron radiation is
given by
\begin{eqnarray}
\left( \frac{{\rm d}^2N_\gamma}{{\rm d}E_\gamma {\rm d}t}\right)_{\rm
  syn} = \frac{\sqrt{3}e^3B(r)\sin\varphi(r)}
     {m_ec^2h}\frac{1}{E_\gamma}G(x),
\end{eqnarray}
where $\varphi(r)$ is the pitch angle at a distance of $r$ on an open
field line, $h$ is the Planck constant, $G(x)=x\int _x^\infty
K_{\frac{5}{3}}(\xi)d\xi$, $K_{\frac{5}{3}}$ is the modified Bessel
function with an order of $5/3$, $x=E_\gamma/E_{\rm syn,\,c}(r)$ and
\begin{eqnarray}
E_{\rm syn,\,c}(r) &&= 1.5h\gamma^2\nu_{\rm L}\sin\varphi(r) \nonumber
\\ && = 4.3\times10^6hB(r)\gamma^2\sin\varphi(r) \label{eq:Esyn}
\end{eqnarray}
is the critical synchrotron photon energy.
The pitch angle $\varphi(r)$ of relativistic primary particles flowing
along a magnetic field line could be small, but it cannot be neglected
for {\SR}. While the pitch angle $\varphi(r)$ of pairs increases due to
the cyclotron resonant absorption of the low-energy photons
\citep{2008ApJ...680.1378H} and it varies with the emission
altitudes. The mean pitch angles of the two types of pairs are
different owing to the effect of cyclotron resonant absorption for
different particles with different Lorentz factors. The synchrotron
radiation from pairs play an important role in X-ray band to soft
$\gamma$-ray band (e.g. $\lesssim 0.02$~GeV).

The photon number spectrum of the curvature radiation can be given by
\begin{eqnarray}
\left( \frac{{\rm d}^2N_\gamma}{{\rm d}E_\gamma {\rm d}t}\right)_{\rm
  cur} = \frac{\sqrt{3}e^2\gamma}{h\rho(r) E_\gamma}G(x),
\end{eqnarray}
where $\rho(r)$ is the curvature radius at $r$ and
Equation~(\ref{Ecr}) shows the critical curvature photon energy.
%

\cite{1970RvMP...42..237B} presented an analytic formula for the
photon spectrum of the inverse Compton scattered photons per electron
in the case of extreme Nishia-Klein limit and then
\cite{2008ApJ...676..562T} gave a simplified form, i.e.,
\begin{eqnarray}
&& \left(\frac{{\rm d}^2N_\gamma} {{\rm d}E_\gamma {\rm
      d}t}\right)_{\rm ICS}=\int _{\varepsilon _ 1}^{\varepsilon _
    2}\frac{3\sigma _{\rm T} c}{4\gamma^2} \cdot \frac{n_{\rm
      syn}(\varepsilon ,r)+n_{\rm X}(\varepsilon ,r)}{\varepsilon}
  \nonumber\\ &&\times\left[2q\ln q+(1+2q)(1-q)+\frac{(\Gamma
      q)^2(1-q)}{2(1+\Gamma q)}\right]{\rm
    d}\varepsilon, \label{eq:qq}
\end{eqnarray}
where $q=E_1/\Gamma(1-E_1)$, $\Gamma=4\gamma\varepsilon/m_ec^2$ and
$E_1=E_\gamma/E_e$. $\varepsilon$ is the energy of soft photons for
scattering, and $\varepsilon_1$ and $\varepsilon_2$ are the minimum
and maximum energy of the soft photons for integration, respectively.
We choose the values of $\varepsilon_1$ and $\varepsilon_2$ to fulfill
the condition of $E_1 < 1$.  The lower limit $\varepsilon_1$ in the
Equation(\ref{eq:qq}) is chosen to be around 1\,eV, and the upper
limit $\varepsilon_2$ is adjusted artificially to make a quick
convergence of the Equation(\ref{eq:qq}).

There are two possible sources of soft photons for the inverse Compton
scattering, one is the thermal photons, and the other is synchrotron
photons. The thermal photons, generated by the stellar surface with a
typical surface temperature $T$, is an important source for Compton
scattering of the {\pps} and {\sps}. The number density of soft
photons is given by
\begin{equation}
n_{\rm X}(\varepsilon,r)=\frac{1}{\pi^2(\hbar
  c)^3}\frac{\varepsilon^2}{\exp(\varepsilon/kT)-1}
\left(\frac{R}{r}\right)^2, \label{n-thermal}
\end{equation}
where $k$ is the Boltzmann constant. The temperature $T$ is taken to
be $2\times10^6\, \rm K$ in our calculations.

\begin{table*}[tb]
\centering
\caption {Best fit parameters for the calculated phase-averaged
  spectrum of the Crab pulsar
\label{tbl_2}}
\begin{tabular}{ccccccccccc}
\hline \hline 
$\alpha$ & $\zeta$ & $\gamma_{\rm min,\,pair}^{\rm CR}$
& $\gamma_{\rm max,\,pair}^{\rm CR}$  & $\varphi_{\rm pair}^{\rm CR}$ & 
$\gamma_{\rm min,\,pair}^{\rm ICS}$
& $\gamma_{\rm max,\,pair}^{\rm ICS}$  & $\varphi_{\rm pair}^{\rm ICS}$ &
$\gamma_{\rm min}^{\rm pri}$ & $\gamma_{\rm max}^{\rm pri}$ &
$\varphi_{\rm pri}$ \\
\hline
$45^{\circ}$ & $63^{\circ}$ & $2.5\times 10^2$ & $2.18\times 10^4$ &
0.0092 & $6.0 \times10^2$ & $1.19\times10^5$ & 0.0074 & $1.0\times
10^6 $ & $2.78\times 10^7$ & 0.000098 \\ \hline
\end{tabular}   %
%
\end{table*}

\begin{figure}
\centering
\includegraphics[angle=0,scale=.58]{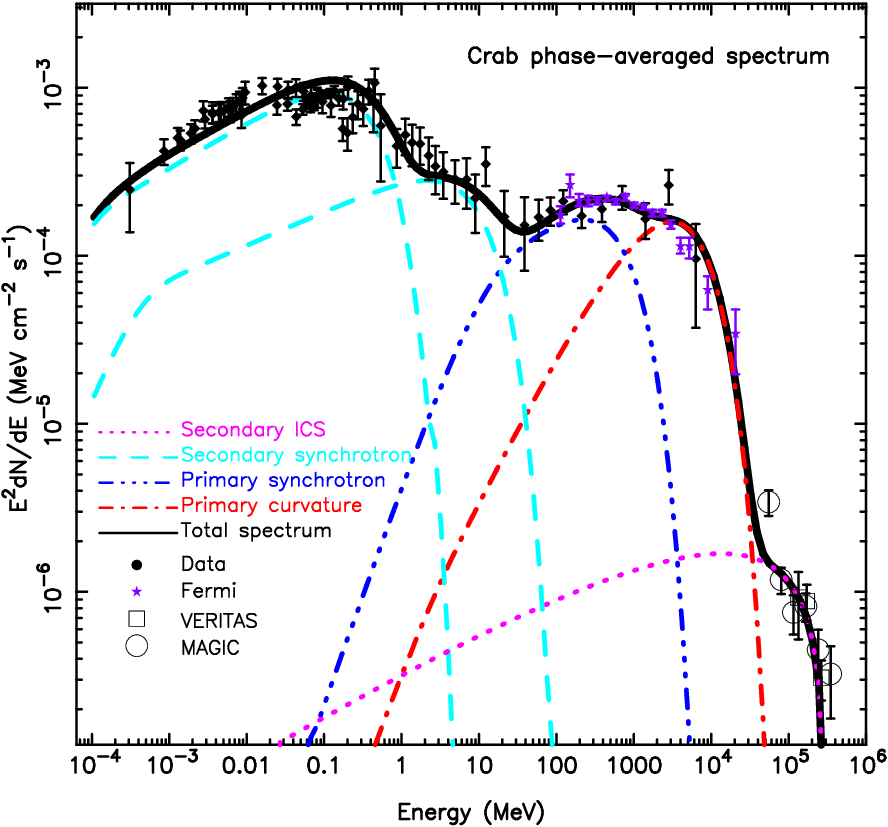}
\caption{The modeled phase-averaged spectrum for the Crab pulsar. The
  calculated total spectrum (thick black solid line) is obtained from
  the sum of {\SR} and ICS from two kinds of pairs and {\CR} and {\SR}
  from {\pps}. It is found that the {\CR} and {\SR} from {\pps} is
  mainly contributed to {\gr} band (20\,MeV to 20\,GeV); {\SR} from
  CR-induced pairs and ICS-induced pairs dominates the X-ray band and
  soft {\gr} band (100\,eV to 10\,MeV). ICS from the pairs contributes
  to hard TeV {\gr} band ($\sim 20$\,GeV to 400\,GeV). The data (solid
  circles) are taken from \cite{kuiper01}. The {\fermi} {\gr} data is
  plotted in purple circles. The {\gr} spectral data of VERITAS
  (square) are taken from \cite{veritas}, and the MAGIC data (big
  hollow circle) are taken from \cite{magic12}. \\ (A color version of
  this figure is available in the online journal.)} \label{PAS}
\end{figure}


\begin{table*}[bht]
\centering
\caption {Best fit parameters for the calculated phase-resolved
  spectra of the Crab pulsar
\label{tbl_3}}
\begin{tabular}{lcccccccccc}
\hline \hline 
Phase band & $\gamma_{\rm min,\,pair}^{\rm CR}$
& $\gamma_{\rm max,\,pair}^{\rm CR}$  & $\varphi_{\rm pair}^{\rm CR}$ & 
$\gamma_{\rm min,\,pair}^{\rm ICS}$
& $\gamma_{\rm max,\,pair}^{\rm ICS}$  & $\varphi_{\rm pair}^{\rm ICS}$ &
$\gamma_{\rm min}^{\rm pri}$ & $\gamma_{\rm max}^{\rm pri}$ &
$\varphi_{\rm pri}$  & $\Delta\Omega_{\rm eff}$\\
\hline
LW1 & $2.5\times 10^2$ & $1.2\times 10^4$ &
0.009 & --- & --- & --- & $1.0\times
10^6 $ & $1.76\times 10^7$ & 0.00005  & 1.82  \\ 
P1 & $2.5\times 10^2$ & $0.9\times 10^4$ &
0.0078 & $4.6 \times10^2$ & $0.86\times10^5$ & 0.0071 & $1.0\times
10^6 $ & $2.47\times 10^7$ & 0.00009  & 3.03 \\ 
TW1 & $2.5\times 10^2$ & $0.95\times 10^4$ &
0.009 & $3.0 \times10^2$ & $0.68\times10^5$ & 0.0079 & $1.0\times
10^6 $ & $2.78\times 10^7$ & 0.00008  &  3.16 \\ 
Bridge & $3.0\times 10^2$ & $0.71\times 10^4$ &
0.0085 & $5.0 \times10^2$ & $0.55\times10^5$ & 0.0070 & $1.0\times
10^6 $ & $2.55\times 10^7$ & 0.00003  & 2.53 \\ 
LW2 & $3.5\times 10^2$ & $1.25\times 10^4$ &
0.008 & $5.0 \times10^2$ & $0.52\times10^5$ & 0.0079 & $1.0\times
10^6 $ & $2.50\times 10^7$ & 0.00005  &  1.78 \\ 
P2 & $5.0\times 10^2$ & $1.22\times 10^4$ &
0.0078 & $5.0 \times10^2$ & $0.82\times10^5$ & 0.0085 & $1.0\times
10^6 $ & $2.97\times 10^7$ & 0.000072  &  3.81 \\ 
TW2 & $5.0\times 10^2$ & $1.3\times 10^4$ &
0.008 & --- & --- & --- & $1.0\times
10^6 $ & $1.82\times 10^7$ & 0.00007  & 2.78  \\ 
\hline
\end{tabular}   %
%
\end{table*}

The other source of soft photons arises from synchrotron radiation of
pairs. Owing to the quite abundant soft synchrotron photons, the
scattering of this kind of photons is more significant at higher
altitudes near the light cylinder.
Although the {\SR} spectrum from a single particles with a Lorentz
factor of $\gamma$ is maintained to a wide energy band (for example,
soft X-ray band to $\gamma$-band), it is more likely to be a spectral
line in fact, which is concentrated on the critical energy. Thus we
can rewrite the synchrotron spectral power using the total energy loss
rate, i.e.,
\begin{equation}
P_{\rm \nu,\,syn} \simeq \dot{E}_{\rm
  syn}\delta(\varepsilon-\varepsilon_{\rm
  syn,\,c}) \simeq \frac{4}{3}\sigma_{\rm T}cU_{\rm
  B}\gamma^2\delta(\varepsilon-\varepsilon_{\rm c}),
\label{simnsyn}
\end{equation}
where $\sigma_{\rm T}$ is the Thompson scattering cross section,
$U_{\rm B}$ is the energy density of the local magnetic field, and
$\delta(\varepsilon-\varepsilon_{\rm c})$ is a Delta function.

\begin{figure*}[!htb]
\centering
\includegraphics[angle=0,scale=.38]{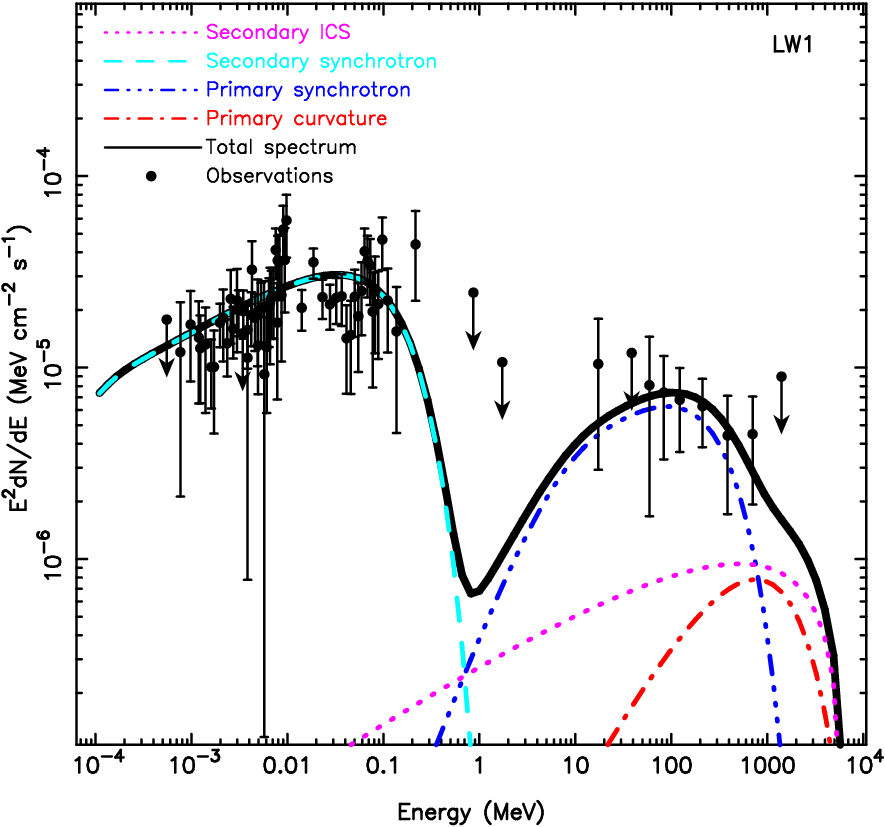}
\includegraphics[angle=0,scale=.38]{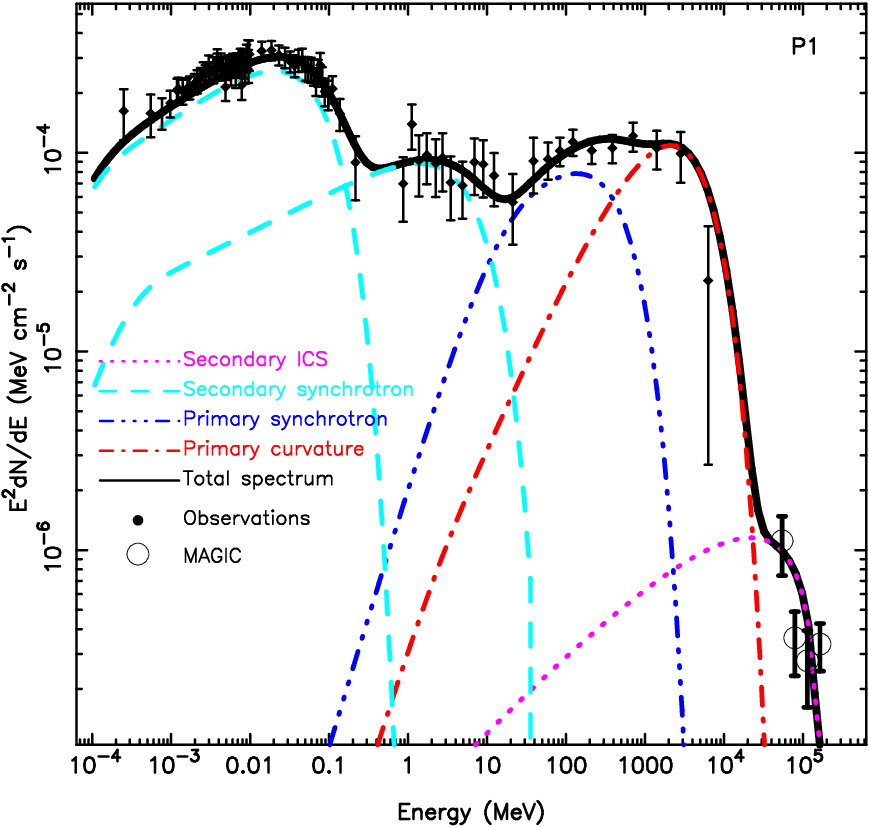}
\includegraphics[angle=0,scale=.38]{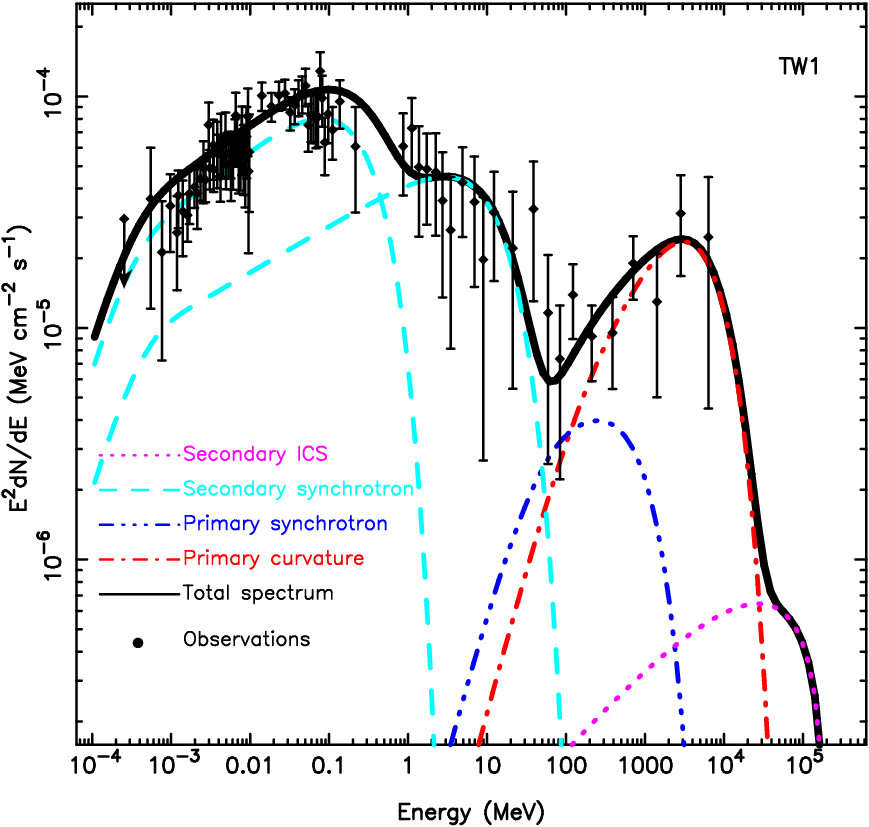}
\includegraphics[angle=0,scale=.38]{f7c.eps}
\includegraphics[angle=0,scale=.38]{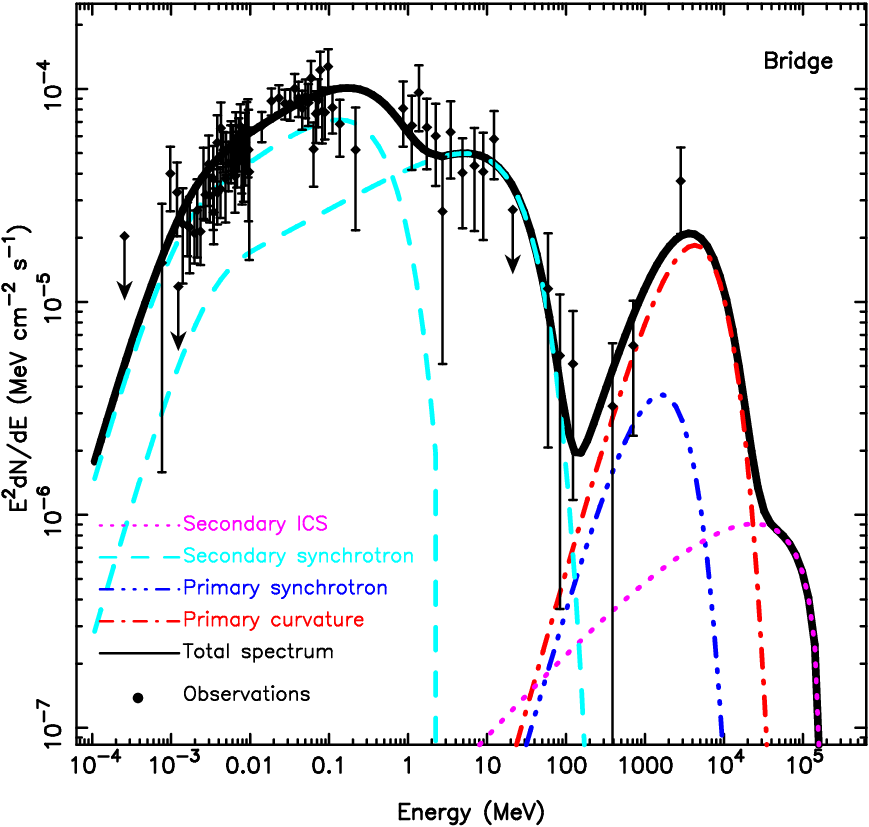}
\includegraphics[angle=0,scale=.38]{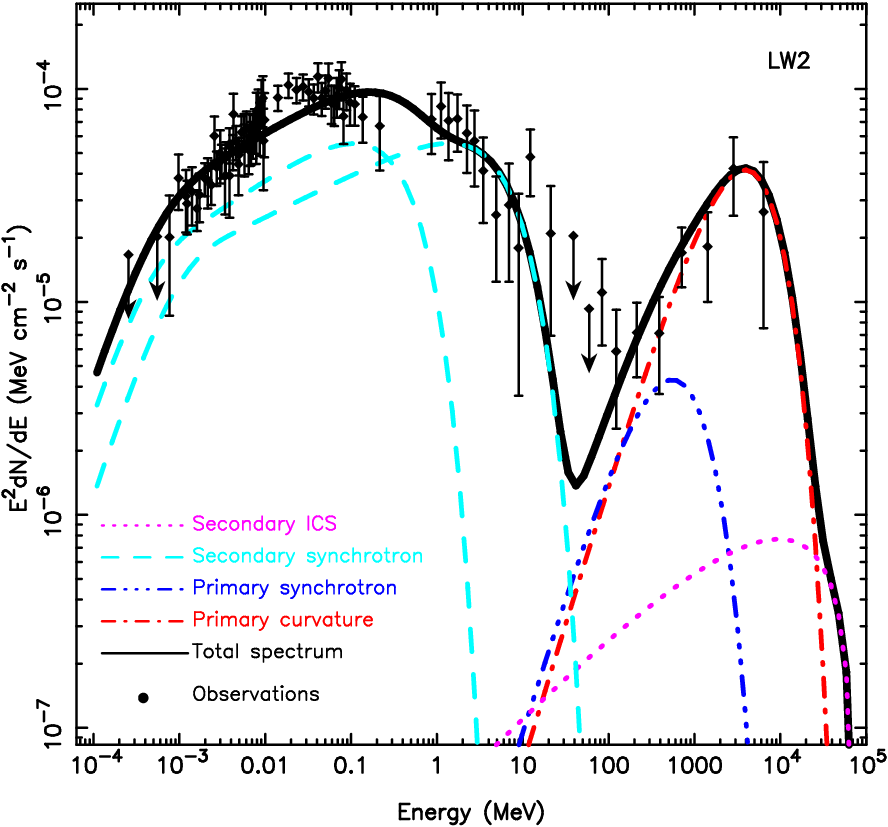}
\includegraphics[angle=0,scale=.38]{f7e.eps}
\includegraphics[angle=0,scale=.38]{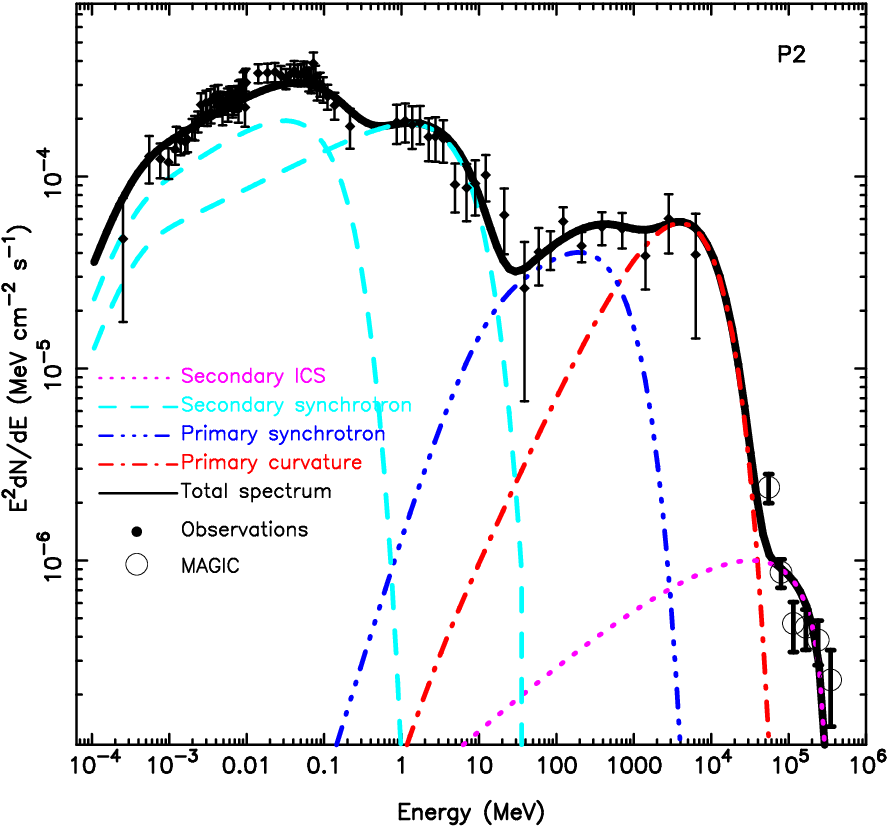}
\includegraphics[angle=0,scale=.38]{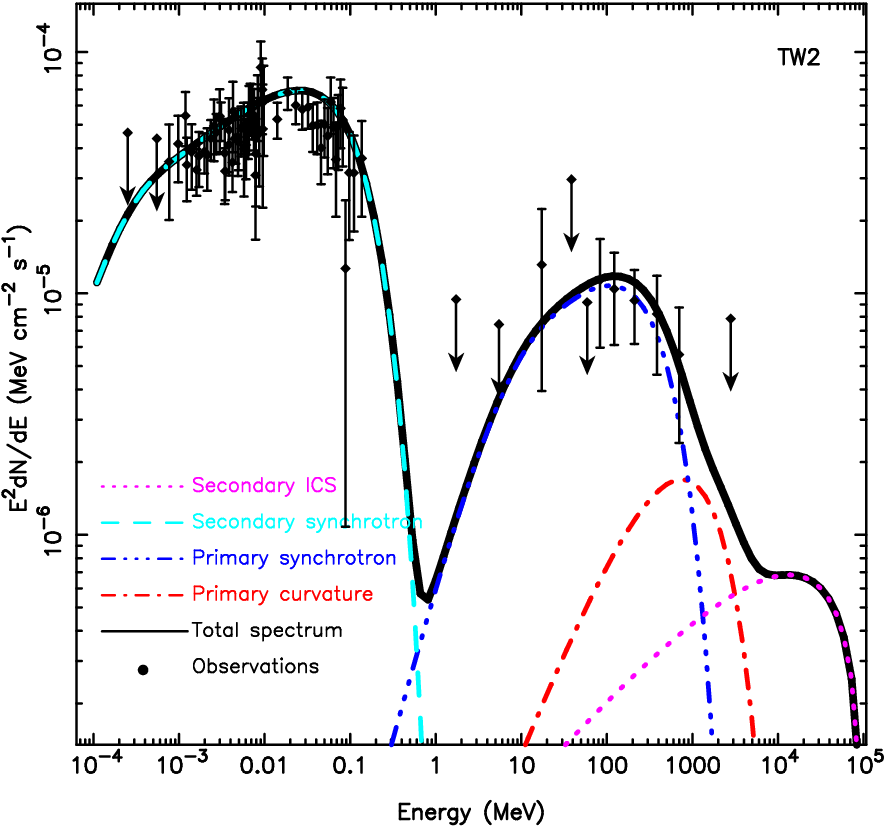}
\caption{Similar as Figure \ref{PAS}, but for modeled seven phase-band
  phase-resolved spectra of the Crab pulsar. TW1 and LW2 are plotted
  twice. The MAGIC (50$-$400\,GeV) spectral data are available only
  for P1 and P2, which are taken from \cite{magic12}, whereas the {\sl
    Fermi} and VERITAS data are not included here.\\ (A color version
  of this figure is available in the online journal.)} \label{PRS}
\end{figure*}

Supposing pairs follow a power-law spectrum $n_{\rm
  e}(\gamma)=n_0\gamma^{-s}$ with a particle energy spectral index of
$s$, we can obtain the synchrotron emissivity $j_\nu$ in a simple form
\begin{equation}
j_\nu = \frac{1}{\Delta \Omega}\int P_{\rm \nu,\,syn} n_{\rm
  e}(\gamma) {\rm d}\gamma = \frac{2\sigma_{\rm T}cU_{\rm
    B}n_0}{3\Delta \Omega \nu_{\rm L}}\left( \frac{\nu}{\nu_{\rm
    L}}\right)^\frac{1-s}{2},
\label{j_syn}
\end{equation}
where $\nu_{\rm L}$ is the Larmor frequency of an electron in the
magnetic field, $n_0$ is a constant and $\Delta\Omega(r)$ is the solid
angle of the beam of synchrotron photons and is estimated as
\begin{equation}
\Delta\Omega(r)=\int _0^{2\pi}\int _0^{\varphi(r)}d\phi\sin\theta d\theta
\approx\pi\varphi^2(r).\label{eq:omega}
\end{equation}
Therefore we use an approximate formula to facilitate ICS calculations
when we consider the soft seed synchrotron photons. The number density
of the soft synchrotron photons, $n_{\rm syn}(\varepsilon ,r)$ can be
given by
\begin{eqnarray}
n_{\rm syn}(\varepsilon ,r) = \frac{\Delta \Omega
  r}{hc\varepsilon}j_\nu = \frac{2\sigma_{\rm T}U_{\rm B} n_0
  r}{3\varepsilon_{\rm L}^2}\left( \frac{\varepsilon}{\varepsilon_{\rm
    L}} \right)^{-\frac{1+s}{2}},
\label{eq:nsyn}
\end{eqnarray}
where $\varepsilon_{\rm L}=h\nu_{\rm L}$ is the Larmor energy.

Following the method of \cite{Du11}, we divide the annular gap region
into 40 rings and 360 equal intervals in the magnetic azimuth, i.e. in
total 40$\times$360 small magnetic tubes. A small magnetic tube has an
area $A_0$ on the neutron star surface. From Equation (\ref{rho}), the
number density of primary particles at a altitude $r$ is $n(r) =
\frac{\Omega B(r)}{2 \pi c\; e} \cos\zeta_{\rm out}$. The
cross-section area of the magnetic tube at $r$ is
$A(r)=B_0A_0/B(r)$. The flowing particle number at $r$ in the magnetic
tube is
\begin{equation}
\Delta N(r) = 
A(r)\Delta s\frac{\Omega B(r)}{2 \pi c e}\cos\zeta_{\rm  out}, 
\label{dnf}
\end{equation}
here $\Delta s$ is the arc length along the field.
The energy spectrum $N_{\rm pri}={\rm d} N/{\rm d} \gamma$ of the
accelerated primary particles is not well understood in the first
physical principle.  Here we assume the primary particles in the
magnetic tube to follow a power-law energy distribution ${\rm d}N/{\rm
  d}\gamma = N_{\rm 0}\gamma^{\Gamma}$ with an index of
$\Gamma=-2.2$. Using Equation (\ref{dnf}), $N_{\rm 0}$ can be derived
by integration the equation above.

\citet{2008ApJ...680.1378H} have assumed a broken power-law
distribution for pairs with indexes of $-2.0$ and $-2.8$ [see their
  Equation (47)]. But in our model, we postulate that the two types of
relativistic pairs follow two different power-law energy distribution
as noted in \S\,2.3, i.e.
\begin{eqnarray}
N_{\rm pairs}(\gamma) = \Bigg\{ \begin{array}{lr} C_1
  \gamma^{-s_1},~\gamma_{\rm min}^{\rm CR} \leqslant \gamma \leqslant
  \gamma_{\rm max}^{\rm CR}, \\ 
\\
C_2 \gamma^{-s_2},~\gamma_{\rm min}^{\rm ICS} \leqslant \gamma \leqslant
\gamma_{\rm \,max}^{\rm ICS}, 
\end{array} \nonumber \label{npair}
\end{eqnarray}
where $s_1=2.45$ and $s_2=2.6$ are the spectral index, $C_1$ and $C_2$
are two coefficients, $\gamma_{\rm min}^{\rm CR}$, $\gamma_{\rm
  max}^{\rm CR}$, $\gamma_{\rm min}^{\rm ICS}$ and $\gamma_{\rm
  max}^{\rm ICS}$ are lower limits and upper limits of the Lorentz
factors for the CR and ICS pairs.

The pairs can be generated with a large multiplicity ($10^3 - 10^5$)
via the $\gamma-B$ process in the lower regions of the annular gap and
the core gap near the neutron star surface. The pitch angle of pairs
increases due to the cyclotron resonant absorption of the low-energy
photons. The mean pitch angle of secondary particles is considerable
owing to the effect of cyclotron resonant absorption. The synchrotron
radiation from secondary particles has some contributions to the
low-energy $\gamma$-ray emission, e.g. $\lesssim 0.02$~GeV.

We also had an analytical formula of optical depth $\tau_{\rm
  \gamma-B}$ due to the $\gamma-$B absorption \citep{LD10}
\begin{equation}
\tau_{\rm \gamma-B}\,(r) = \frac{1.55\times10^7 r}{E_{\rm
    \gamma}}K_{1/3}^2 (\frac{2.76\times10^6 r^{5/2}P^{1/2}}{B_{\rm
    0,\,12}R^3 E_{\rm \gamma}} ),
\label{tau}
\end{equation}
here $E_{\rm \gamma}$ is in units of MeV, $B_{\rm 0,\,12}$ is the
surface magnetic field in units of $10^{12}$\,G. We found that the
photons of the Crab pulsar with an energy of $<100$\,GeV always
satisfy the condition of $\tau_{\rm \gamma-B}\ll 1$ if the emission
altitude is greater than a few hundred kilometers.
Thus the final multi-wavelength spectrum emitted by the {\pps} and
{\sps} can be calculated by
\begin{eqnarray}
F(E_\gamma) & = & \frac{1}{\Delta \Omega_{\rm eff} D^2}\int_{\gamma_{\rm
    min}}^{\gamma_{\rm max}}\Bigg[ \left( \frac{ {\rm d}^2 N_\gamma}
  {{\rm d}E_\gamma {\rm d}t }\right)_{\rm cur} + \left( \frac{{\rm
      d}^2 N_\gamma} {{\rm d}E_\gamma {\rm d}t }\right)_{\rm syn}
   \nonumber \\ & + & \left( \frac{{\rm d}^2 N_\gamma}{{\rm d}E_\gamma
    {\rm d}t }\right)_{\rm ICS}\Bigg] e^{-\tau_{\rm \gamma-B}\,(r)},
\label{fflux}
\end{eqnarray}
where $\Delta\Omega_{\rm eff}$ is the effective solid angle of the
emission beam, and $D=2$\,kpc is the distance from the Crab pulsar to
the Earth.

There are actually six main spectral components for the pulsar total
spectrum. Based on our calculations, the synchrotron and curvature
radiation from primary particles and the synchrotron radiation and ICS
from secondary particles are required to calculate the phase-averaged
and phase-resolved spectra, while ICS from {\pps} and CR from pairs
can be ignored for the Crab pulsar. When we calculate the SR spectra,
we carefully deal with the modified Bessel function with for
unprecedent accuracies. While calculating ICS spectrum, as stated
above, we use Equations (\ref{eq:qq}) and (\ref{n-thermal}) and
(\ref{eq:nsyn}) to reduce the computation time.

To accelerate our computations, we use the method of ``averaged
emission-altitude" \citep[as introduced in][]{Du11} to calculate the
four contributive spectral components for the Crab pulsar, i.e.,
synchrotron and curvature radiation from primary particles, and ICS
and {\SR} from pairs. We first obtain the emission altitudes from
Figure (\ref{EMH}) for each phase band. For instance, the emission
altitude is about $0.54\,R_{\rm LC}$ on the field line of a magnetic
azimuth $\psi=-101^\circ$ for P1; while for P2, the emission altitude
is $0.49\,R_{\rm LC}$ on the field line of $\psi=109^\circ$. Then we
compute their acceleration {\ef} $E_\parallel$ and potential $\Psi$,
and adjust the minimum and maximum Lorentz factor for both primary
particles and pairs ($\gamma_{\rm min}^{\rm pri}$ and $\gamma_{\rm
  max}^{\rm pri}$ and $\gamma_{\rm min}^{\rm 2nd}$ and $\gamma_{\rm
  max}^{\rm 2nd}$), the pitch angle $\varepsilon(r)$ and the
$\gamma$-ray beam angle $\Delta\Omega_{\rm eff}$ to fit the
multi-wavelength {\pas} and {\prs} for the Crab pulsar.

We fitted the phase-averaged spectrum and seven phase-band
phase-resolved spectra of the Crab pulsar, and the results are shown
in Figure~\ref{PAS} and Figure~\ref{PRS}, respectively. For {\pas}, we
basically used the multi-wavelength data from \cite{kuiper01} and
combined with latest {\fermi}, MAGIC and VERITAS {\gr} spectral data,
whereas for {\prs}, we only used the multi-wavelength data from
\cite{kuiper01}.
The best fit parameters for phase-averaged spectrum and phase-resolved
spectra are listed in Table \ref{tbl_2} and Table \ref{tbl_3},
respectively. From spectra fitting, we found that
the calculated {\gr} spectra are not sensitive to $\gamma_{\rm
  min}^{\rm pri}$, but quite sensitive to $\gamma_{\rm max}^{\rm pri}$ 
which is chosen below the value obtained from the balance of curvature
radiation and radiation reaction shown in Figure~\ref{DYN}. The solid
angle $\Delta\Omega_{\rm eff}$ was always assumed to be 1 by many
authors for simplicity. We adjusted it as a free parameter around 1
for different phases.

We found that multi-wavelength emission from the phase bands of P1, P2
and bridge contribute significantly to the total {\pas}. The {\pas}
and {\prs} are decomposed into four spectral components.  Curvature
radiation and {\SR} from primary particles is the main origin of the
observed $\gamma$-ray emission (10\,MeV to $\sim$ 20\,GeV), {\SR} from
CR-induced pairs and ICS-induced pairs dominates the X-ray band and
soft {\gr} band. ICS from the pairs contributes significantly up to
TeV band, while ICS from both {\pps} and {\CR} from pairs can be
neglected for the spectrum fitting. Owing to the larger emission
altitudes (which leads to lower acceleration {\ef}) for LW1 and TW2,
the lorentz factors $\gamma$ are very low which results from lower
acceleration {\ef}, {\CR} and {\SR} have little contributions to the
LW1 and TW2 phase band spectra. The TeV emission of ICS from pairs can
be also found for P1, bridge and P2 in our calculated {\prs}, which
are consistent with the modeled TeV light curve.
Two types of pairs, CR-induced and ICS-induced, could be therefore
confirmed by the spectra of the Crab pulsar, these may be also the
origin of the two types of wind pairs in Crab Nebula.

\section{Conclusions and Discussions}

Owing to its strong multi-wavelength emission, the famous Crab pulsar
is a crucial astrophysical object to distinguish the emission
mechanisms from different magnetospheric models. In this paper, we
calculated radio, X-ray, {\gr} and TeV light curves, phase-averaged
spectrum and phase-resolved spectra in the framework of the 3D annular
gap and core gap model with reasonable emission-geometry parameters
($\alpha=45^\circ$ and $\zeta=63^\circ$).
It is found that the {\ef} in the {\AG} is huge ($\sim 10^{16}$\,eV)
in the several tens of neutron star radii and vanishes near the region
of $R_{\rm LC}$. The {\pps} are accelerated to ultra-relativistic
energies, and produce numerous {\sps} (CR and ICS pairs) in the inner
region of the {\AG} via $\gamma-$B process.
The pulsed emission of radio, X-ray and $\gamma$-ray are generated
from the emission of {\pps} or {\sps} (pairs) with different emission
mechanisms in the nearly similar region of the {\AG} (or {\CG}) in
only one pole, this leads to the ``phase-aligned'' multi-wavelength
{\lcs}. The emission of P1 and P2 is originated from the annular gap
region near the null charge surface, while the emission of bridge is
mainly originated from the core gap region. 
%

%

Assuming that power-law energy distributions of {\pps} and two types
of pairs, the {\pas} and {\prs} of the Crab pulsar are well produced
by mainly four components: {\SR} from CR-induced and ICS-induced pairs
dominates the X-ray band to soft {\gr} band (100\,eV to 10\,MeV);
{\CR} and {\SR} from the {\pps} mainly contribute to {\gr} band
(10\,MeV to $\sim$ 20\,GeV); ICS from the pairs significantly
contributes to the TeV {\gr} band (100\,GeV to 400\,GeV). {Our fitted
  {\pas} and {\prs} have similar tendency varying with the photon
  energy and are basically consistent with outer gap model
  \citep{2008ApJ...676..562T} and the slot gap model
  \citep{2008ApJ...680.1378H} at soft X-ray to a few tens of GeV band,
  but quite different in $>20$\,GeV band. This is mainly due to the
  additional spectral component of ICS from pairs. From the
  multi-wavelength spectral fitting we emphasize that {\CR} alone
  emitted from the {\pps} cannot explain the TeV band (25 to 400 GeV)
  emission of the {\pas}, ICS from pairs significantly contribute to
  this $>20$\,GeV band. In addition, two types of pairs are generated
  in the magnetosphere, and they may be also the origin of the two
  types of wind pairs in Crab Nebula \citep{fermi-Crab}.

Radio emission (1.4\,GHz) of the Crab pulsar is originated from a
narrow and high-altitude region with a similar location of the
$\gamma$-ray emission, which leads to the phase-aligned peaks. Our
model for radio photon sky-map is patch-like, however the detailed
emission mechanism for radio emission is needed to further
studied. 

The popular outer gap (except the versions of
\cite{1995ApJ...438..314R,2000ApJ...537..964C,Romani10}) and slot gap
models are two-pole models
\citep{2008ApJ...676..562T,2009ApJ...707L.169Z,2010ApJ...725.2225L,
  2008ApJ...680.1378H}. To model the observed {\lcs} and spectra for
the Crab pulsar, they require the emission from the both magnetic
poles, which result from larger magnetic inclination angle $\alpha$ or
larger viewing angle $\zeta$.
However, our {\AG} model is an intermediate emission-altitude and
single-pole model with reasonable $\alpha$ and $\zeta$ from the X-ray
torus fitting \citep{NG08}.
Unfortunately, the important parameters $\alpha$ and $\zeta$ for
pulsar emission geometry are uncertain so far. \cite{NG08} can only
give a reliable viewing angle $\zeta$ for some young pulsars which
have X-ray torus configurations, and then combine with the radio
rotating vector model (RVM) to obtain the inclination angle $\alpha$
using the radio polarization angle (PA) fitting. The simple RVM model
is only based on the geometry at a certain low altitude for an assumed
circular emission beam, and the propagation effects that can change
the polarization states that had been already ignored by the RVM
model. The derived $\alpha$ by this method is therefore debatable. A
better method is strongly desired to obtain the convincing values of
$\alpha$ and $\zeta$.
 
To well explain the multi-wavelength pulsed emission from pulsars, the
detailed magnetic field configuration and 3D global acceleration
electric field distribution with proper boundary conditions for the
annular gap and the core gap should be carefully studied. 
Unfortunately, these two physical aspects are not fully
understood. Recnetly, \cite{Romani10} studied pulsar light curves with
magnetosphere beaming models and found that outer gap model and
approximating force-free dipole field were preferred at their high
statistical significance. However, \cite{Harding11} also studied
high-energy pulse profiles (e.g. the Vela pulsar) using both retarded
vacuum dipole and force-free field geometry. They found that slot gap
model with vacuum dipole was more favorable. Therefore, the subject of
pulsar magnetic field configuration is still debatable.
In addition, the problem of 3D acceleration field with the general
relativistic effect and pair screening effect is more complicated,
although many efforts have been paid. We just derived the 1D (actually
2D) continuous solution for the acceleration Possion equation, and the
general relativistic and pair screening effects have not been taken
into account in our annular gap model at present. This is our first
step to establish our model picture, and will benefit further 3D
complicated physical studies with considerations of related
effects. We emphasize that some simplified hypothesis considering
qualitative physical effects have been used in our model to study
pulsar light curves and spectra. This can give us insightful
enlightenments to improve our knowledge of pulsar radiation physics.
We will further improve our model to give more precise modeled {\lcs},
especially for the phases of LW1 and TW2.

In sum, the multi-wavelength emission from the Crab pulsar can be well
explained in the annular gap and core gap model, and this is also done
for the Vela pulsar \citep{Du11}. Our model is a promising model to
unveil the multi-wavelength pulsed emission from {\gr} pulsars.

\acknowledgments 
The authors are very grateful to the referee for the insightful and
constructive comments. We thank both the pulsar groups of NAOC and of
Peking University for useful conversations. The authors are supported
by NSFC (10821061, 10573002, 10778611, 10773016, 11073030 and
10833003) and the Key Grant Project of Chinese Ministry of Education
(305001). 

{\sl Facilities: Fermi} (LAT), MAGIC, VERITAS

\end {document}